\documentclass[9pt,conference]{IEEEtran} 
\IEEEoverridecommandlockouts

\usepackage{cite}
\usepackage{amsmath,amssymb,amsfonts}
\usepackage{algorithmic}
\usepackage{graphicx}
\usepackage{textcomp}
\usepackage[utf8]{inputenc}
\usepackage{multirow}
\usepackage{booktabs}
\usepackage{colortbl}
\usepackage[table]{xcolor}
\usepackage{arydshln}
\begin{document}
\renewcommand{\baselinestretch}{0.956}\normalsize

\title{Adapter-Based Multi-Agent AVSR Extension for Pre-Trained ASR Models}

\author{\IEEEauthorblockN{Christopher Simic}
\IEEEauthorblockA{
\textit{Technische Hochschule Nuernberg}\\
Nuernberg, Germany \\
christopher.simic@th-nuernberg.de}
\and
\IEEEauthorblockN{Korbinian Riedhammer}
\IEEEauthorblockA{
\textit{Technische Hochschule Nuernberg}\\
Nuernberg, Germany \\
korbinian.riedhammer@th-nuernberg.de}
\and
\IEEEauthorblockN{Tobias Bocklet}
\IEEEauthorblockA{
\textit{Technische Hochschule Nuernberg}\\
Nuernberg, Germany \\
tobias.bocklet@th-nuernberg.de}
}

\maketitle

\begin{abstract}

We present an approach to Audio-Visual Speech Recognition that builds on a pre-trained Whisper model. 
To infuse visual information into this audio-only model, we extend it with an AV fusion module and LoRa adapters, one of the most up-to-date adapter approaches.
One advantage of adapter-based approaches, is that only a relatively small number of parameters are trained, while the basic model remains unchanged. Common AVSR approaches train single models to handle several noise categories and noise levels simultaneously. 
Taking advantage of the lightweight nature of adapter approaches, we train noise-scenario-specific adapter-sets, each covering individual noise-categories or a specific noise-level range. The most suitable adapter-set is selected by previously classifying the noise-scenario. This enables our models to achieve an optimum coverage across different noise-categories and noise-levels, while training only a minimum number of parameters. 

Compared to a full fine-tuning approach with SOTA performance our models achieve almost comparable results over the majority of the tested noise-categories and noise-levels, with up to 88.5\% less trainable parameters. Our approach can be extended by further noise-specific adapter-sets to cover additional noise scenarios. It is also possible to utilize the underlying powerful ASR model when no visual information is available, as it remains unchanged.
\end{abstract}


\section{Introduction}

Automatic Speech Recognition (ASR) and Audio-Visual Speech Recognition (AVSR) share common targets, as both aim to transcribe input sequences. ASR approaches rely on audio-only speech data, which can be particularly challenging in noisy environments, resulting in increased Word Error Rates (WER).
AVSR incorporates additional visual information, typically recordings of lip movements, which is one way to achieve a WER reduction, especially in noisy environments.

Common AVSR approaches train models from scratch on audio-visual data. In contrast, our approach builds on a pre-trained ASR model to benefit from the modeling capabilities of models that are trained on huge amounts of speech data. In this work, we utilize the SOTA ASR model Whisper, which is currently the most robust model for speech-to-text transcription. In general, our approach can be applied to any pre-trained ASR model. 

To adapt the pre-trained audio-only ASR model to the additional visual modality, we use LoRa adapters instead of full fine-tuning the model. These adapters allow a baseline model to be adapted to a specific domain without changing the underlying baseline model's basic behavior. These adapters can be switched on and off in case no visual information is available or changing environmental conditions. 

Conventional AVSR approaches train single models to handle simultaneously all noise categories and a wide range of noise levels. Due to the very lightweight adapters, several domain-specific adapter-sets can be kept available at all times and switched dynamically. We take advantage of this by training multiple specific adapter-sets that enable customized adaptation to specific noise scenarios. Additionally, we include a simple classification layer that determines the noise scenario based on the input audio signal to select the most suitable adapter-set for processing.

Our key contributions are: 
\begin{itemize} 
\item The first approach to integrate adapters into a pre-trained ASR model to enable processing of additional visual information, with a detailed analysis and comparison across multiple noise types and levels
\item The first multi-agent system for Audio-Visual Speech Recognition by developing multiple adapter-sets to optimize the performance across various noise categories and levels
\item Implementing noise category or level classification to dynamically select the most suitable adapter-set for each scenario
\end{itemize}
\section{Related Work}

\subsection{ASR and AVSR}

Automatic Speech Recognition (ASR) and Audio-Visual Speech Recognition (AVSR) approaches target the automatic transcription of input signals (audio-only or audio-visual).
Conformer and transformer architectures have established in these domains.
Beside conformer approaches~\cite{ASR_conformer_2020,ASR_conformer2_2022} espacially transformer approaches like Whisper~\cite{ASR_whisper_2022}, HuBERT~\cite{ASR_HuBERT_2021}, and Wav2Vec2~\cite{ASR_wav2vec_2020} have gained popularity in ASR. Whisper can be emphasized here as it exhibits a comparatively high robustness in moderate noise conditions~\cite{ASR_whisperAT_2023}. However, all ASR approaches suffer from increased Word Error Rates (WER) under noisy conditions.

One way, to tackle this problem is the development of transformer~\cite{AVSR_avhubert_2022, AVSR_robust_ssAVSR_avhubert2_2022, AVSR_jointly_rawD_2022, AVSR_AV_FTof_aOnlyASR} and conformer-based\cite{AVSR_autoAVSR_2023, AVSR_e2e_conformer,AVSR_AVFormer} AVSR approaches, which use visual information from lip movements to reduce the WER, espacially for noisy conditions.
The best performance for AVSR is currently achieved by the approaches RAVEn~\cite{AVSR_jointly_rawD_2022}, AV-HuBERT~\cite{AVSR_robust_ssAVSR_avhubert2_2022} and AUTO-AVSR~\cite{AVSR_autoAVSR_2023}, meaning that all of them can be considered as state-of-the-art (SOTA). 
While most common AVSR approaches train models from scratch, Whisper-Flamingo~\cite{AVSR_2024__WhisperFlamingo} and our previous work AV Fusion~\cite{AVSR_2023_selfsupervised} use pre-trained Whisper models. AV-Fusion combines the audio-visual features in a fusion module located upstream of the encoder. Whisper-Flamingo works with separate audio and visual encoders and fuses the modalities in the decoder.

\subsection{Adapter Approaches}
Adapter approaches like LoRa~\cite{adapt_lora_2022} and AdaLoRa~\cite{adapt_adalora_2023} allow transformer and conformer fine-tuning, without changing the baseline model’s parameters. For this purpose, both insert additional blocks, consisting of two serial linear layers alongside each linear layer within the baseline model’s attention blocks. The rank, which describes the dimension between both linear layer is chosen to be small in order to significantly reduce the number of trainable parameters compared to a full fine-tuning. 
Various approaches demonstrate that adapters are suitable for adapting pre-trained ASR models to changed domains~\cite{adapt_whisperChildSp_2024, adapt_SURE_2023, adapt_peft_ser_2023, adapt_distilWh_2024, adapt_resTransf_2024, AVSR_AVFormer} or to extend Large Language Models (LLM) with additional modalities~\cite{mllm_oct_2024, mllm_tunLN_2024}. The advantage of trainable adapters, apart from the reduced number of parameters to be trained, is their ability to be used as a lightweight add-on to the baseline model and falling back to the baseline model's basic behaviour if adapters are switched off.

\begin{figure}[t]
  \centering
  \includegraphics[width=0.8\linewidth]{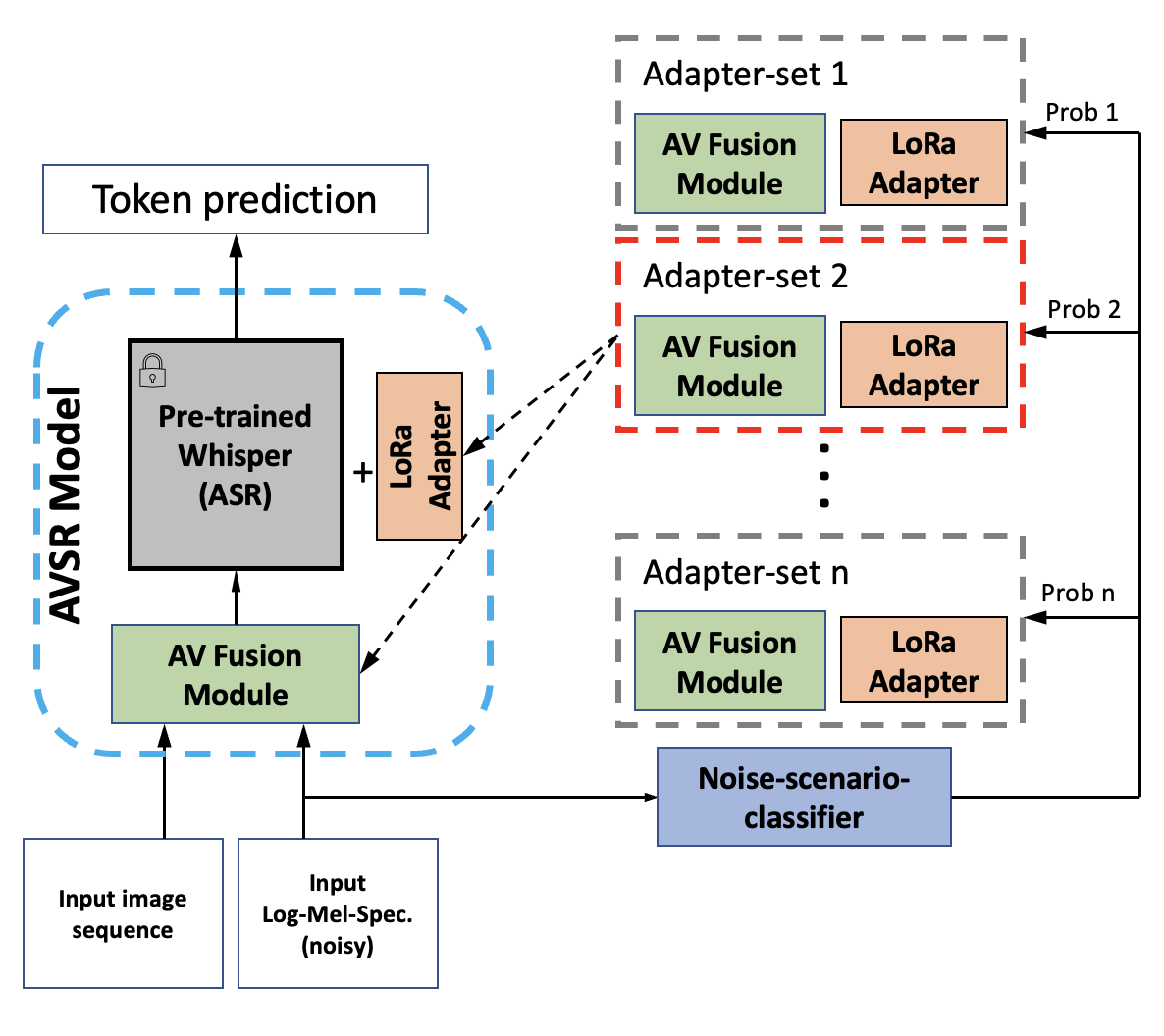}
  \caption{Overall model architecture. AVSR model on the left, including the frozen, pre-trained ASR model (gray) and the selected set of LoRa adapters (orange) and the AV fusion module (green). Noise-scenario-classifier (blue) to select the most suitable adapter-set on the right.}
  \label{fig:overview_model}
\end{figure}
\section{Data}

\subsection{LRS3}

We utilize the LRS3-TED dataset~\cite{DATA_2018_LRS3}, the most comprehensive and challenging dataset for AVSR, to train and evaluate our models. LRS3 contains 150k sequences from TED Talks on YouTube, featuring over 9k speakers and a vocabulary of 51k words. The dataset comprises more than 430 hours of video, divided into Pretrain (407 hours), Trainval (30 hours), and Test (1 hour), which we adopt for our experiments.

\subsection{VoxCeleb2}

VoxCeleb2~\cite{DATA_2018_Voxcel2} is an audio-visual dataset originally designed for speaker recognition, without provided transcriptions. As our approach is trained self-supervised, we can utilize this dataset. It contains over one million utterances from more than 6k speakers. While the dataset is multilingual, we focus solely on the English sequences with approximately 1700 hours.

\subsection{Musan}

To systematically contaminate speech data with synthetic noise, we employ the Musan dataset~\cite{DATA_2015_musan}, which provides diverse audio samples across different categories. 60 hours of speech, 42.5 hours of music, and 6 hours of natural sounds. Following the AV-HuBERT~\cite{AVSR_avhubert_2022} methodology, we generate babble noise by merging 30 speech samples. We split each noise category into 80\% for training, 10\% for validation, and 10\% for testing to prevent any overlap.
\section{Method}

\subsection{Overall Model}

Figure~\ref{fig:overview_model} shows our model's overall structure, with the AVSR model on the left and multiple adapter-sets and the noise-scenario-classifier on the right. 

The AVSR model consist of three parts, a pre-trained, frozen ASR model (gray), a group of LoRa adapter (orange) and an AV fusion module (green). Each adapter-set on the right consists of a spicific set of LoRa adapters and an AV fusion module, which are kept to replace the adapter-set in the AVSR model if necessary. The classifier (blue) receives the noisy input mel-spectrum to determine the current noise scenario and select the optimum adapter-set.

Like \cite{AVSR_2023_selfsupervised, AVSR_2024__WhisperFlamingo}, we choose Whisper as the core element of our AVSR model, which is currently one of the most powerful ASR models. As Whisper can only process audio information, the model must be extended by the upstream AV fusion module, which is adopted from~\cite{AVSR_2023_selfsupervised}. This fusion module consists two processing levels. The first level includes two separate CNN-based feature extraction modules for audio and visual inputs. The extracted audio and visual feature vectors are processed by a multi-layer multi-head cross-attention module, which produces the inputs for the Whisper-based model. Further details about the fusion module architecture can be taken from~\cite{AVSR_2023_selfsupervised}. 

In contrast to~\cite{AVSR_2023_selfsupervised} and~\cite{AVSR_2024__WhisperFlamingo}, which perform a full fine-tuning for the pre-trained ASR model, we add LoRa adapters to the ASR model. These adapters provide the advantage that the ASR model weights are kept frozen while only the adapter weights are trained, which means that the base model's basic behavior remains unchanged. We insert adapters to all linear layers of the ASR transformer model's Query, Value, Key and Output layers with a rank of 64. 
Initially, we also tested AdaLoRa adapters. These perform an SVD to determine the optimum rank for each layer, under the constraint that the average rank across all layers remains constant. Against our expectations, it turned out that LoRa adapters yield better results for the tested parameters.

Usually AVSR models aim to simultaneously cover all noise categories and the entire Signal-to-Noise Ratio (SNR) range. We benefit from the lightweight nature of adapter-based approaches and train several noise-scenario-specific adapter-sets that share the same base model. Each adapter-set covers only a small part of the entire noise spectrum.
Due to their lightweight character, the adapter-sets can be easily exchanged when noise scenarios change. We distinguish between two basic concepts and analyze which concept provides better results:
\begin{itemize}
\item Noise-category-specific -- One specific adapter-set for each of the noise categories Babble, Music, Noise and Single Sidespeaker
\item Noise-level-specific -- One adapter-set each for the SNR range above or below 0dB
\end{itemize}

Like ~\cite{AVSR_2023_selfsupervised} we use the frozen Whisper ASR model to generate target values from clean audio inputs at the model levels mel-spectrum level, embedding level after the encoder and logit level after the decoder.
The AVSR model receives synthetically contaminated audio inputs and corresponding video frames, that contain the speaker's lip movements.
During training, we calculate a loss for all three model levels, which are added to a weighted total loss. 

\begin{equation}
\text{L}_{\text{pre}} = 0.5 \cdot L1\left(Mel_c,Mel_n\right) + L1\left(Emb_c,Emb_n\right)
\end{equation}

\begin{equation}
\text{L}_{\text{ft}} = \text{L}_{\text{pre}} + CE\left(Dec_c,Dec_n\right)
\end{equation}

$\text{L}_{\text{pre}}$ defines the loss used during pre-training. Therefore, we calculate the  Mean Absolute Error (MAE) between the model receiving clean inputs and the trained AVSR model receiving noisy inputs at the mel-spectrum and embedding level after the encoder. The mel-spectrum loss is semi-weighted to reduce its influence on the training process.
This loss aims to prevent the generated mel-spectrum from drifting too far from natural mel-spectrums.
$\text{L}_{\text{ft}}$ defines the loss during the fine-tuning. This extends $\text{L}_{\text{pre}}$ by the loss term at the logit level after the decoder. We follow the example of the original Whisper training and use a Cross-Entropy loss at this level.

\subsection{Training Details}

We use two model sizes for our approach, based on the Whisper variants base (74M) and small (244M). The applied fusion module contains 13.2M parameters. The LoRa adapters with rank 64 contain 4.8M parameters for base and 14.3M for small. For the noise-category-specific concept we use four adapter-sets and fusion modules that share one Whisper model. This results in a total of 146M model parameters, of which 72M are trainable for the base version and a total of 354M model parameters with 110M trainable parameters for the larger small version. For the noise-level-specific concept, only two adapter-sets and fusion modules share one Whisper model. This results in 110M total model parameters with 36M trainable parameters for the smaller version and 299M total model parameters with 55M trainable parameters for the larger version.

Our models are trained with a batch size of 16. For the first training step, we use the Voxceleb2 dataset for 112k iterations to pre-condition the models. This is followed by 112k iterations with the LRS3 dataset. For these two steps, we only train fusion module and encoder adapters with a learning rate of 1e-4. Afterwards, we train the the fusion module and all adapters using $\text{L}_{\text{ft}}$. For this, we also use the LRS3 dataset and start with a learning rate of 1e-5 for 21k iterations. This is followed by 21k iterations with the learning rate gradually reduced to 1e-7. Due to convergence problems, the learning rate for the larger small models was reduced by 0.5 after the first training step.
We follows the suggestion of~\cite{AVSR_2023_selfsupervised} and choose a SNR range from -15dB to 30dB during training.

\subsection{Adapter-Set Selector}

As we train individual adapter-sets for different noise scenarios, the noise category or noise level must be determined before processing. For fully automatic adapter-set selection, we built a classifier which is illustrated as the blue block in Figure~\ref{fig:overview_model} on the right. 
For this purpose, we utilize a 10-layer CNN ResNet with 64 processing channels. To determine the noise category, the CNN ResNet is followed by two fully connected linear layer and one Softmax layer. This results in four classes representing the four noise categories. For training, we use a Cross-Entropy Loss.
To determine the noise level, which is defined by the Signal-to-Noise Ratio (SNR) value, the CNN ResNet is also followed by two fully connected linear layers, with a single output neuron in the second layer. Since we predict continuous SNR values, we use a Mean Squared Error Loss (MSE) during training.

\section{Results}

\begin{table*}[ht]
\footnotesize
\centering
\scalebox{0.75}{
\begin{tabular}{c|cc|cccc|cccc|cccc|cccc}
\multicolumn{1}{c}{} & \multicolumn{1}{c}{} & \multicolumn{1}{c}{} & \multicolumn{4}{c}{\textbf{Babble SNR [dB]}} & \multicolumn{4}{c}{\textbf{Music SNR [dB]}} & \multicolumn{4}{c}{\textbf{Natural SNR [dB]}} & \multicolumn{4}{c}{\textbf{Sidesp. SNR [dB]}} \\
\cmidrule(rl){4-7} \cmidrule(rl){8-11} \cmidrule(rl){12-15} \cmidrule(rl){16-19}
\textbf{Models} & \textbf{TrP} & \textbf{ToP} & {-10} & {0} & {10} & {20} & {-10} & {0} & {10} & {20} & {-10} & {0} & {10} & {20} & {-10} & {0} & {10} & {20} \\
\midrule

AV-Fusion base FFT~\cite{AVSR_2023_selfsupervised} & 
87M & 87M & 46.7 & 7.1 & 2.9 & 2.5 & 9.4 & 3.2 & 2.7 & 2.4 & 14.2 & 4.1 & 3.0 & 2.5 & 9.9 & 5.2 & 3.0 & 2.5 \\

\\[-0.6em]
\hline
\\[-0.6em]

(Ours) LoRa-AVSR base - Full noise spectrum & 
18M & 92M & 60.7 & 10.0 & 2.9 & 2.5 & 13.7 & 4.1 & 2.8 & 2.5 & 18.8 & 4.7 & 2.7 & 2.4 & 16.6 & 8.6 & 3.7 & 2.4 \\

\\[-0.6em]
\hline
\\[-0.6em]

(Ours) LoRa-AVSR base - Noise category spec. (Babble) & 
18M & 92M & \cellcolor{gray!20}51.5 & \cellcolor{gray!20}8.3 & \cellcolor{gray!20}2.8 & \cellcolor{gray!20}2.6 & 48.5 & 7.9 & 3.3 & 2.8 & 36.1 & 8.3 & 3.3 & 2.7 & 122.7 & 79.1 & 6.6 & 2.8 \\

(Ours) LoRa-AVSR base - Noise category spec. (Music) & 
18M & 92M & 77.1 & 13.4 & 3.0 & 2.6 & \cellcolor{gray!20}11.4 & \cellcolor{gray!20}3.8 & \cellcolor{gray!20}3.0 & \cellcolor{gray!20}2.7 & 19.2 & 5.1 & 2.8 & 2.6 & 87.4 & 44.3 & 5.6 & 2.8 \\

(Ours) LoRa-AVSR base - Noise category spec. (Noise) & 
18M & 92M & 80.0 & 14.4 & 3.2 & 2.8 & 21.9 & 5.1 & 3.2 & 2.9 & \cellcolor{gray!20}17.3 & \cellcolor{gray!20}4.4 & \cellcolor{gray!20}2.9 & \cellcolor{gray!20}2.6 & 124.2 & 82.7 & 7.6 & 2.9 \\

(Ours) LoRa-AVSR base - Noise category spec. (Sidespeaker) & 
18M & 92M & 90.8 & 15.3 & 3.4 & 2.6 & 22.1 & 5.2 & 3.0 & 2.6 & 26.6 & 6.0 & 3.0 & 2.6 & \cellcolor{gray!20}8.7 & \cellcolor{gray!20}5.1 & \cellcolor{gray!20}2.9 & \cellcolor{gray!20}2.5 \\

\\[-0.85em]
\hdashline
\\[-0.85em]

(Ours) LoRa-AVSR base - Noise category spec. + Classifier  & 
72M** & 146M** & \cellcolor{green!15}51.6 & \cellcolor{green!15}8.3 & \cellcolor{green!15}2.8 & \cellcolor{green!15}2.6 & \cellcolor{green!15}11.5 & \cellcolor{green!15}3.9 & \cellcolor{green!15}3.0 & \cellcolor{green!15}2.7 & \cellcolor{green!15}17.2 & \cellcolor{green!15}4.4 & \cellcolor{green!15}2.9 & \cellcolor{green!15}2.5 & \cellcolor{green!15}9.2 & \cellcolor{green!15}5.3 & \cellcolor{green!15}2.9 & \cellcolor{green!15}2.6 \\

\\[-0.6em]
\hline
\\[-0.6em]

(Ours) LoRa-AVSR base - Noise level spec. (HighNoise) & 
18M & 92M & \cellcolor{gray!20}48.7 & \cellcolor{gray!20}7.8 & 3.2 & 2.7 & \cellcolor{gray!20}9.6 & \cellcolor{gray!20}3.5 & 3.1 & 2.9 & \cellcolor{gray!20}14.5 & \cellcolor{gray!20}4.3 & 2.7 & 2.8 & \cellcolor{gray!20}7.7 & \cellcolor{gray!20}6.2 & 10.7 & 6.8 \\

(Ours) LoRa-AVSR base - Noise level spec. (LowNoise) & 
18M & 92M & 78.4 & 10.5 & \cellcolor{gray!20}3.0 & \cellcolor{gray!20}2.6 & 20.1 & 4.5 & \cellcolor{gray!20}3.0 & \cellcolor{gray!20}2.7 & 22.7 & 5.2 & \cellcolor{gray!20}3.0 & \cellcolor{gray!20}2.6 & 45.9 & 10.9 & \cellcolor{gray!20}3.3 & \cellcolor{gray!20}2.7 \\
\\[-0.85em]
\hdashline
\\[-0.85em]
(Ours) LoRa-AVSR base - Noise level spec. + Classifier & 
36M** & 110M** & \cellcolor{blue!15}48.8 & \cellcolor{blue!15}7.8 & \cellcolor{blue!15}3.0 & \cellcolor{blue!15}2.6 & \cellcolor{blue!15}9.6 & \cellcolor{blue!15}3.6 & \cellcolor{blue!15}3.0 & \cellcolor{blue!15}2.7 & \cellcolor{blue!15}14.6 & \cellcolor{blue!15}4.4 & \cellcolor{blue!15}3.0 & \cellcolor{blue!15}2.6 & \cellcolor{blue!15}10.9 & \cellcolor{blue!15}6.7 & \cellcolor{blue!15}5.4 & \cellcolor{blue!15}2.7 \\
\\[-0.5em]
\hline
\hline
\\[-0.5em]


AV-HuBERT large*~\cite{AVSR_robust_ssAVSR_avhubert2_2022} & 
477M & 477M & 34.9 & 5.8 & 2.0 & -- & 9.7 & 2.5 & 1.8 & -- & 9.7 & 2.5 & 1.8 & -- & 11.4 & 2.9 & 1.8 & -- \\
\\[-0.6em]
\hline
\\[-0.6em]
AV-Fusion small FFT~\cite{AVSR_2023_selfsupervised}& 
257M & 257M & 38.7 & 4.7 & 2.2 & 1.9 & 6.7 & 2.4 & 1.9 & 1.9 & 10.6 & 2.8 & 2.1 & 2.0 & 5.4 & 2.8 & 2.0 & 1.9 \\

\\[-0.6em]
\hline
\\[-0.6em]
(Ours) LoRa-AVSR small - Full noise spectrum & 
28M & 272M & 52.0 & 6.1 & 2.1 & 1.9 & 9.1 & 2.6 & 2.0 & 1.9 & 12.7 & 3.2 & 2.1 & 1.9 & 8.9 & 4.1 & 2.2 & 1.9\\

\\[-0.6em]
\hline
\\[-0.6em]

(Ours) LoRa-AVSR small - Noise category spec. (Babble) & 
28M & 272M & \cellcolor{gray!20}46.2 & \cellcolor{gray!20}6.2 & \cellcolor{gray!20}1.9 & \cellcolor{gray!20}1.9 & 38.5 & 4.4 & 2.2 & 2.0 & 23.8 & 5.0 & 2.1 & 1.8 & 119.8 & 49.8 & 3.0 & 1.9 \\

(Ours) LoRa-AVSR small - Noise category spec. (Music) & 
28M & 272M & 67.3 & 8.0 & 2.2 & 1.9 & \cellcolor{gray!20}8.9 & \cellcolor{gray!20}2.5 & \cellcolor{gray!20}1.9 & \cellcolor{gray!20}2.0 & 14.7 & 3.4 & 2.1 & 1.9 & 82.6 & 27.6 & 2.8 & 2.0 \\

(Ours) LoRa-AVSR small - Noise category spec. (Noise) & 
28M & 272M & 71.8 & 9.0 & 2.1 & 1.8 & 15.4 & 3.1 & 2.0 & 1.9 & \cellcolor{gray!20}13.0 & \cellcolor{gray!20}3.1 & \cellcolor{gray!20}2.0 & \cellcolor{gray!20}1.9 & 117.3 & 52.1 & 3.0 & 1.9 \\

(Ours) LoRa-AVSR small - Noise category spec. (Sidespeaker) & 
28M & 272M & 74.2 & 9.2 & 2.2 & 1.9 & 15.7 & 3.1 & 2.2 & 1.9 & 18.2 & 3.7 & 2.2 & 1.9 & \cellcolor{gray!20}6.5 & \cellcolor{gray!20}3.7 & \cellcolor{gray!20}2.0 & \cellcolor{gray!20}1.8\\
\\[-0.85em]
\hdashline
\\[-0.85em]
(Ours) LoRa-AVSR small - Noise category spec. + Classifier & 
110M** & 354M** & \cellcolor{green!15}46.2 & \cellcolor{green!15}6.3 & \cellcolor{green!15}1.9 & \cellcolor{green!15}1.9 & \cellcolor{green!15}8.9 & \cellcolor{green!15}2.6 & \cellcolor{green!15}1.9 & \cellcolor{green!15}2.0 & \cellcolor{green!15}13.0 & \cellcolor{green!15}3.1 & \cellcolor{green!15}2.0 & \cellcolor{green!15}1.9 & \cellcolor{green!15}7.0 & \cellcolor{green!15}3.8 & \cellcolor{green!15}2.1 & \cellcolor{green!15}1.8\\

\\[-0.6em]
\hline
\\[-0.6em]

(Ours) LoRa-AVSR small - Noise level spec. (LowNoise) & 
28M & 272M & \cellcolor{gray!20}44.6 & \cellcolor{gray!20}6.4 & 2.3 & 2.0 & \cellcolor{gray!20}7.0 & \cellcolor{gray!20}2.5 & 2.1 & 2.1 & \cellcolor{gray!20}11.0 & \cellcolor{gray!20}3.2 & 2.1 & 1.9 & \cellcolor{gray!20}5.6 & \cellcolor{gray!20}3.9 & 5.3 & 3.4 \\
(Ours) LoRa-AVSR small - Noise level spec. (HighNoise) & 
28M & 272M & 75.1 & 8.0 & \cellcolor{gray!20}2.1 & \cellcolor{gray!20}1.8 & 14.9 & 2.9 & \cellcolor{gray!20}2.0 & \cellcolor{gray!20}1.9 & 18.0 & 4.3 & \cellcolor{gray!20}2.0 & \cellcolor{gray!20}1.8 & 40.8 & 6.8 & \cellcolor{gray!20}2.1 & \cellcolor{gray!20}1.8\\

\\[-0.85em]
\hdashline
\\[-0.85em]

(Ours) LoRa-AVSR small - Noise level spec. + Classifier & 
55M** & 299M** & \cellcolor{blue!15}44.6 & \cellcolor{blue!15}6.4 & \cellcolor{blue!15}2.1 & \cellcolor{blue!15}1.8 & \cellcolor{blue!15}7.1 & \cellcolor{blue!15}2.5 & \cellcolor{blue!15}2.0 & \cellcolor{blue!15}1.9 & \cellcolor{blue!15}11.1 & \cellcolor{blue!15}3.2 & \cellcolor{blue!15}2.0 & \cellcolor{blue!15}1.8 & \cellcolor{blue!15}8.4 & \cellcolor{blue!15}4.3 & \cellcolor{blue!15}3.3 & \cellcolor{blue!15}1.8 \\

\\[-0.5em]
\bottomrule
\\[-0.5em]

\end{tabular}
}
\caption{
WER [\%] for AV-HuBERT large, AV-Fusion (with full fine-tuning) and several of ours adapter-based model combinations. The models are grouped by model size. TrP indicates the number of trainable parameters, ToP the number of total model parameters. WER values are provided for different noise categories and noise levels. * Values for AV-HuBERT are taken from the official paper.~\cite{AVSR_robust_ssAVSR_avhubert2_2022} ** For all classifier-based models, all available adapter-set parameters are listed, while only one particular adapter-set is used for inference (same number of parameters as for the noise-specific models listed above)}
\label{tab:res}
\end{table*}

\tablename~\ref{tab:res} provides the results for our models and two benchmark models, AV-HuBERT~\cite{AVSR_robust_ssAVSR_avhubert2_2022} and AV-Fusion~\cite{AVSR_2023_selfsupervised}. All models are divided into two groups according to their model size, with the smaller models in the upper part and the larger models in the lower part. All models are tested on the noise categories Babble, Music, Natural and Sidespeaker across a wide SNR range from -10dB to 20dB, with -10dB representing scenarios with very high background noise levels.
All results are given in Word Error Rate (WER) in [\%].

\subsection{AV-HuBERT vs. AV-Fusion}

To ensure comparability, we trained the AV-Fusion~\cite{AVSR_2023_selfsupervised} models, from the official repo for both model sizes on our data. This was necessary because AV-Fusion does not provide models that were pre-trained on VoxCeleb2 and fine-tuned on 430h LRS3. 
Since AV-HuBERT does not provide a seed definition, our splits are not identical and a test on our data would produce different results.
Consequently, we use the values from the official paper~\cite{AVSR_robust_ssAVSR_avhubert2_2022}. We are limited to the large version, as no values are available for the smaller base version. In contrast to AV-HuBERT, we analyze Music and Natural noise separately, so we report the mean values given by AV-HuBERT for both noise categories.

Comparing the larger AV-Fusion small FFT model to AV-HuBERT large reveals that AV-HuBERT is superior for Babble noise, while AV-Fusion has clear advantages for Sidespeaker noise. We also see a mixed picture for Music and Natural noise. 
Across all SNR values, that AV-HuBERT provides results for (-10dB, -5dB, 0dB, 5dB, 10dB) and all noise categories AV-Fusion achieves an average reduction in WER of 2.1\%, which demonstrates that both approaches have a very similar performance, with a slight advantage for AV-Fusion.
Another AVSR with similar performance is Whisper-Flamingo~\cite{AVSR_2024__WhisperFlamingo}. The official paper only provides one value for SNR 0dB and the noise category Babble with a WER of 5.7, which is also slightly higher than the value of 4.7 achieved by AV-Fusion.

Since all of these approaches show quite similar performance, we will focus on the comparison to AV-Fusion FFT in the subsequent discussion, as results for both model sizes and SNR values from -10 dB to 20 dB are available for this approach.

\subsection{LoRa Adapter}

The LoRa-AVSR base/small - Full noise spectrum models share the same model architecture as all subsequent noise-scenario-specific models, but have been trained simultaneously on all four noise categories and the entire SNR range.

Compared to the AV-Fusion FFT models, which were trained on the same data, the results reveal an average increase of 41.5\% for high noise levels (-10dB) across all noise categories. For the SNR range from 0dB to 20dB, the average increase in WER is reduced to 11.0\%. 
For these models, the number of trainable parameters is 79.3\% to 89.1\% lower than for AV-Fusion and 94.1\% lower than for AV-HuBERT large, which explains the lower performance.

\subsection{Noise-Scenario-Specific Adapter-Sets}

\tablename~\ref{tab:res} provides the results for noise-category and noise-level-specific adapter models for both model sizes and a comparison to AV-Fusion with full fine-tuning and AV-HuBERT large.
The noise-category-spcific adapters were trained on individual noise categories spanning the entire SNR range from -15dB to 30dB. The noise-level-specific adapters were trained on the SNR range greater than or less than 0dB, but received examples from all noise categories. All adapter-sets share the same frozen Whisper models.
The results for the noise scenarios each model was trained for, are highlighted in gray, and show that noise-scenario-specific training leads to significant improvements compared to the LoRa-AVSR - Full noise spectrum models. The noise-level-specific adapter-set models in particular show a clear improvement compared to training over the entire SNR range, demonstrating the benefits of using adapter-sets for specific noise scenarios. 
The WER values are still worse, but relatively close to those of AV-Fusion FFT, which is remarkable since AV-Fusion trains many times more parameters. For the smaller models and the noise category Sidespeaker, we actually achieve an improvement compared to AV-Fusion for an SNR value of -10db.

In general, it is noticeable that the noise-level-specific models perform slightly better than the noise-category-specific models. Especially for high noise levels at -10dB and 0dB, the noise-level-specific models achieve lower WER values across both model sizes and the majority of noise categories. We hypothesize that the noise-level-specific HighNoise models are forced to pay high attention to the visual information during training, as the audio inputs are very noisy. In contrast, the noise-category-specific models receive many examples from the SNR range between 5dB to 30dB and 5\% clean audio inputs, which may have the effect that these models do not require the visual information to generate correct embedding and logits, for the majority of training samples.

The blue/green marked rows represent the results achieved when the noise-category (green) and noise-level (blue) are not known a-priori but estimated with our proposed classifier for adapter-set selection (sec. IV.C).
The noise-category-classifier achieves a recognition rate of 98.1\%. The noise-level-classifier actually performs a SNR regression. We specify a threshold value of 5dB, which is used for the HighNoise/LowNoise selection, leading to a correct classification rate of 94.7\%. 

All noise-scenario-specific models with classifier achieve results that are almost identical to the optimum results for each noise-scenario-specific model, which demonstrates the strength of this approach. The only exception is the noise category Sidespeaker, especially for the noise-level-specific models.

We identified two reasons for the weakness at this noise category: First, for Sidespeaker noise, the gap between the results for the HighNoise and LowNoise models is larger than for any other noise category, especially at -10dB, which leads to a stronger negative influence of misclassifications. We suspect that this gap could be reduced by presenting a few examples from the LowNoise range to the HighNoise models during training, to get them used to these noise scenarios, and reduce the gap.
Second, the noise-level-classifier performs poorer for this noise category compared to other noise categories. (Noise-Level-Classifier - correct classification rate at SNR -10dB: Babble 100.0\% / Music 98.4\% / Natural 93.3\% / Sidespeaker 91.3\%). An explanation might be that this noise category overlaps two equivalent speech signals and the classifier struggles to identify the main signal, leading to incorrect SNR predictions.

All noise-scenario-specific models achieve a significant improvement for both model sizes compared to the LoRa-AVSR - Full noise spectrum models and the results tend to be close to those of the AV-Fusion FFT, but are still slightly poorer. Comparing the number of trainable parameters reveals that the noise-level-specific models in particular have 58.6\% to 78.6\% less trainable parameters than AV-Fusion and 88.5\% less trainable parameters than AV-HUBERT large.

As part of several ablation studies, we tested the AV-Fusion FFT models' ability to handle audio-only information, to simulate the case that no visual information is available, which led to an dramatic increase in WER values even at low noise levels. This demonstrates one of the main advantages of our adapter-based models. If no visual information is available, the underlying unmodified ASR model can be used, which is very capable on audio-only data as it has been trained on huge amounts of speech data. Using the AV-Fusion small FFT model with 257M parameters, an extra Whisper small model with 244M parameters needs to be provided for this case.

Another advantage is the easy expandability, as the underlying Whisper model remains unchanged. This allows the extension by further adapter-sets, for instance additional noise categories or even more specific adapter-sets such as a Babble noise adapter-set explicitly trained on high noise levels to improve the adaptation to this specific scenario. In case of changing noise categories, a fully fine-tuned AV-Fusion or AV-HuBERT model would require a re-training.

\section{Conclusion}

We presented an adapter-based approach to Audio-Visual Speech Recognition that allows to train noise-scenario-specific adapter-sets to cover a large range of noise levels and different noise categories. 
We deploy two classifiers with very high recognition rates to select the optimum adapter-set for each specific noise scenario.
We demonstrate that the AV-Fusion approach, which is related to our approach, achieves a performance comparable to one of the current SOTA AVSR approaches, AV-HuBERT.
Compared to AV-Fusion, our models reveal a slightly higher word error rate on average, which is partly caused by certain weaknesses towards the noise category Sidespeaker. Across large ranges of tested SNR values and noise categories, our models show a performance close to that of AV-Fusion and AV-HuBERT, despite 88.5\% less trainable parameters and smaller overall model size compared to AV-HuBERT. 

The moderate weaknesses in terms of performance are compensated by the advantages such as the expandability with additional noise scenario-specific adapter-sets and the usability of the unchanged underlying powerful Whisper ASR model, in case no visual information is available.



\end{document}